\documentclass[10pt,letterpaper]{article}%
\usepackage{opex3}
\usepackage{color}
\usepackage{graphicx}
\usepackage{amsmath}%
\usepackage{amsfonts}%
\usepackage{amssymb}
\providecommand{\U}[1]{\protect\rule{.1in}{.1in}}

\begin{document}

\title{Absolute efficiency estimation of photon-number-resolving detectors using twin beams}

\author{A. P. Worsley, H. B. Coldenstrodt-Ronge*, J. S. Lundeen, P. J. Mosley, B. J.
Smith, G. Puentes, N. Thomas-Peter, and I. A. Walmsley}

\address{University of Oxford, Clarendon Laboratory, Parks Road, \\ Oxford,
OX1 3PU, United Kingdom}

\vskip-.3cm \parskip0pc\hskip2.25pc \footnotesize 
   \parbox{.8\textwidth}{\begin{center}\it*Corresponding author: \it\textcolor{blue}{\underline{h.coldenstrodt-ronge1@physics.ox.ac.uk}} \rm \end{center} } \normalsize  \vskip-.2cm

\begin{abstract}
A nonclassical light source is used to demonstrate experimentally the
absolute efficiency calibration of a photon-number-resolving detector.
The photon-pair detector calibration method
developed by Klyshko for single-photon detectors is generalized to take
advantage of the higher dynamic range and additional information provided by photon-number-resolving detectors. This enables the use of brighter twin-beam sources including amplified pulse pumped sources, which increases the relevant signal and provides measurement redundancy, making the calibration more robust.
\end{abstract}
 


\definecolor{orange}{rgb}{1,0.4,0}
\definecolor{purple}{rgb}{0.2,0,1}

\section{Introduction}

Quantum optics enables one to make measurements that are more precise than
the fundamental limits of classical optics \cite{Giovannetti2004}. Central to
this capability are quantum optical detectors, those that are sufficiently
sensitive to discern the inherent discreteness of light. These detectors are
key to emerging quantum technologies such as quantum imaging and lithography,
in which the standard wavelength limit to resolution is surpassed by using quantum states of
light and photon-number sensitivity. However, the majority of quantum optical detectors have a response that saturates at only one photon, imposing a significant limitation on the brightness of the optical fields that can be used for such quantum technologies. The result of this binary detector response is a measurement that can discriminate only between zero photons and one or more photons arriving simultaneously; the detector produces an identical response for any number of photons greater than zero. To allow increased brightness of the light sources used in the technologies outlined above (and the concomitant improvement in accuracy that this brings), it is necessary to use detectors that can discern the number of photons incident simultaneously on the detector --- a photon-number-resolving detector (PNRD). Indeed, the development of PNRDs is an active area of research including photomultiplier tubes, the extension of avalanche photodiodes (APDs) to higher photon number through time multiplexing \cite{Haderka2004,Achilles04} or spatially-multiplexed arrays \cite{Jiang2007},
transition-edge sensors (TES) \cite{Cabrera1998,Miller2003,Rosenberg2005},
charge-integration photon detectors (CIPS) \cite{Fujiwara2006},
superconducting single-photon detectors \cite{Somani2001,Hadfield2005},
visible-light photon counters (VLPCs) \cite{Kim1999}, and quantum dot field
effect transistors \cite{Shields2000,Kardynal2007}.

Calibration of optical detectors is a difficult problem. The standard approach uses a previously calibrated light source. The drawback of this approach is
that errors in the source brightness translate directly into errors in the detector efficiency calibration. The converse is also true, leading to a detector and light-source calibration dilemma. To get
around this, brightness calibration is typically based on a fundamental
physical process, for example, the luminosity of blackbody radiation of gold at its
melting point \cite{Booker1987}, or heating of a cryogenic bolometer
\cite{Gentile1996}. Such methods are suitable for calibrating bright light
sources. In contrast, using this method to calibrate detectors operating at the quantum level (i.e. fields containing only a few photons) requires sources with powers on the order of a femtowatt to avoid saturation. Such sources are impractical.

Nonclassical states of light allow us to circumvent the horns of the dilemma due to their behavior in the presence of loss. Realistic detectors can be modeled by optical loss (i.e. attenuation) followed by unit efficiency detectors. The transformation of classical states (e.g. a coherent state, thermal state, etc.) under loss is
parameterized only by source brightness, which scales linearly with the
loss. In contrast, nonclassical states change their statistical character upon
experiencing loss: correlations in optical phase, intensity, photon number,
and electric field transform in a non-trivial way under loss. Based on this fact, Klyshko proposed a way to calibrate
detectors based on the statistical character of light rather than its brightness
\cite{Klyshko80}. This approach relies on spontaneous
parametric downconversion (SPDC) as a light source in which a photon is simultaneously created
in each of two optical modes, usually denoted signal and idler. Since the
photons are created in pairs the two output modes are perfectly correlated in photon number. Thus detection of the idler without the simultaneous detection of the signal photon can only be
caused by loss in the signal arm. Measuring the detected rate of idler photons
$R_{i}$ and photon pairs $R_{c}$ then allows a calibration of the detector efficiency
$\eta_{s}$
\begin{equation}
\eta_{s}=\frac{R_{c}}{R_{i}}, \label{equ:originalKlyshko}%
\end{equation}
and vice versa with $i\leftrightarrow s$. This efficiency estimation was shown
experimentally in \cite{Rarity87,Penin1991,Kwiat1993,Migdall1996}.

The Klyshko scheme is limited by three factors. First, it relies on the
implicit assumption that at most one photon pair is emitted at a time, a feature which
can be violated if the SPDC is pumped strongly. Indeed more than one pair
is often desirable, as in continuous-variable experiments.
In this case, the detector efficiency can be obtained by first lowering
the pump power and using the Klyshko method \cite{Erlangen08}. However, the detector is  now calibrated
outside the regime of its intended use, which could necessitate
subsequent assumptions such as the independence of the estimated efficiency from the pump power. Thus, the direct \textit{in situ} calibration of detectors would be desirable.
Second, the Klyshko scheme is primarily designed for single-photon detectors and is not directly translated to PNRDs. And third, this calibration method is highly sensitive to the input state quality and measurement uncertainties. Because the number of measurements taken is exactly the number needed to determine
the efficiency, any errors in the measurements will propagate directly into
the efficiency estimation.

Here we present a technique for measuring the absolute quantum efficiency of a detector based on the Klyshko method, but explicitly taking into account multiple photon events. This improves the calibration accuracy and allows for calibration of a PNRD. This approach utilizes the PNRD capability to measure the photon-number distribution of an optical mode. Using two PNRDs the joint photon-number statistics between the two electromagnetic field modes, including photon-number correlations and individual photon-number distributions, of the SPDC source can be determined. For each element of the resulting joint
photon statistics, one can find a formula giving the detector efficiencies of the two PNRDs.
For the zero- and one-click elements, this reproduces the
Klyshko result. However, the increased dynamic range of PNRDs
allows the use of brighter sources that produce measurable rates in the higher
photon-number elements of the joint statistics. We then use optimization techniques to estimate the
detector efficiencies from the increased number of measurements. This added
redundancy improves the tolerance of our scheme to background light and
statistical noise. We experimentally demonstrate this efficiency estimation
method with two time-multiplexed PNRDs \cite{Achilles04}.

We begin by introducing a general treatment of PNRDs and then use this
as a basis for describing our generalized Klyshko method. This is followed by the 
description of an experiment to test the efficiency estimation with PNRDs.

\section{Photon-number-resolving detectors}

\label{genanalysis}
Photon-number-resolving detectors (PNRDs) are
a class of photodetectors that have a unique response for every input
photon-number state within their range. Ideally these responses can be
perfectly discriminated. However, the less than perfect efficiency of
realistic detectors causes these responses to overlap, and thus does not allow
for direct photon-number discrimination. Overlap of detector responses can
also arise from the detector electronics (e.g. amplification) or the
underlying detector design. Despite this overlap, the linear relationship
between the detector response and the input state allows for the
reconstruction of the input photon statistics from the measured outcome
statistics. This linear relationship is encapsulated by
\begin{equation}
P_{n}=\mathrm{Tr}\left[  \hat{\rho}\hat{\Pi}_{n}\right]  ,\label{equ:Trace}%
\end{equation}
where $\hat{\rho}$ is the input-state density matrix, $P_{n}$ is the
probability for the $n$th measurement outcome and $\hat{\Pi}_{n}$ is the associated positive operator-value
measurement (POVM) operator. Consider the matrices expressed in the
photon-number basis. Since PNRDs do not contain an optical phase reference,
the off-diagonal elements of $\hat{\Pi}_{n}$ are zero, meaning the
photon-number-resolving detection is insensitive to off-diagonal elements in
$\hat{\rho}.$ It is thus useful to write the diagonal elements of $\hat{\rho}%
$, the photon-number statistics, as a vector $\vec{\sigma}$. Similarly, we
write the outcome probabilities $\left\{  P_{i}\right\}  $ as a vector
$\vec{p}$. In the following, we truncate $\vec{\sigma}$ at photon number
$N-1,$ where $N$ is the number of detector outcomes, although this is not
strictly necessary.

The POVM operators of a general PNRD can be modeled by dividing
the detector performance into two components: efficiency and
detector design, described by matrices $\mathbf{F}$ and $\mathbf{L}$
respectively. In the photon number basis, a POVM element can
be written as \cite{Coldenstrodt-Ronge2008}
\begin{equation}
\hat{\Pi}_{n}=\sum_{m}\left[  \mathbf{F}\cdot\mathbf{L}\left(  \eta\right)
\right]  _{n,m}\left\vert m\right\rangle \left\langle m\right\vert
,\label{equ:explicitPOVM}%
\end{equation}
where $\left[  \mathbf{F}\cdot\mathbf{L}\left(  \eta\right)  \right]  _{n,m}$
corresponds to the probabilty of detecting $n$ out of $m$ incidence photons.
As pointed out in the introduction, detector efficiency $\eta$
can be modeled by a preceding optical loss $l$ of $\left(  1-\eta\right)  $.
In the context of a PNRD, loss causes the photon-number statistics to
transform according to $\vec{\sigma}\rightarrow\mathbf{L}\left(  \eta\right)
\vec{\sigma}$, where
\begin{equation}
\mathbf{L}_{i,j}\left(  \eta\right)  =
\begin{cases}
\binom{j}{i}\eta^{i}\left(  1-\eta\right)  ^{j-i} & \text{if }j\geqslant i\\
0 & \text{otherwise}%
\end{cases}
.\label{equ:DefL}%
\end{equation}
These matrix elements transform the state by lowering photon numbers from $j$
to $i$, representing the loss of photons through a binomial process with
probability $l=\left(  1-\eta\right)  $.

Although the detector-design component of the model depends on
the detailed functioning of the device, a large class of PNRDs --
mode-multiplexers -- can be treated in the same way. These detectors divide an
input optical field mode into many spatial and/or temporal modes and then use
single-photon detection on each mode to achieve number resolution. Examples
include nanowire superconducting detectors \cite{Somani2001}, VLPCs \cite{Kim1999}, intensified charge-coupled devices (CCDs) \cite{Perina2003}, integrated APD arrays \cite{Jiang2007,Hamamatsu}, and time-multiplexed
detectors (TMD). All of these detectors suffer from detector saturation; the
one-photon detector response occurs if two or more photons occupy the same
mode. This saturation effect is modelled by a detector design matrix
$\mathbf{F} = \mathbf{C}$, the ``convolution"
matrix, which has a general but complicated analytic form given in
\cite{Coldenstrodt-Ronge2008} (in the context of the TMD). The form of
$\mathbf{C}$ depends on relatively few parameters comprising the splitting ratios of the
input mode into each of the multiplexed modes and the total number of these
modes.

As an example, in a CCD array detector the pixel shape and size
defines the detected optical mode. The spatial overlap of the detector mode
of each pixel with the incoming optical mode then gives the corresponding
splitting ratio. The total number of pixels is the total number of multiplexed
modes. As the number of multiplexed modes goes to infinity the $\mathbf{C}$
matrix goes to the identity. All PNR detectors can be described by a POVM,
which can be reconstructed by detector tomography
\cite{Coldenstrodt-Ronge2008,Lundeen2008}. For PNR detectors that do not rely
on mode multiplexing, such as transition-edge superconducting detectors, and
single APD detectors, one must factor the POVM elements into an $\mathbf{F}$ matrix and an $\mathbf{L}$ matrix to apply the
calibration procedure described in this paper.

Focusing on the specific case of mode-multiplexed detectors, one
can rewrite the Eq. (\ref{equ:explicitPOVM}) as,
\begin{equation}
\vec{p}=\mathbf{C}\cdot\mathbf{L}\left(  \eta\right)  \cdot\vec{\sigma
},\label{equ:singleTMD}%
\end{equation}
where the elements of the $i$th row of $\mathbf{C}\cdot\mathbf{L}\left(
\eta\right)  $ are the diagonals of the $i$th operator in the POVM set
$\left\{  \hat{\Pi}_{i}\right\}  $.

With PNRDs in two beams, denoted $1$ and $2$, one can measure not
only the individual photon-number statistics $\vec{\sigma_{1}}$ and
$\vec{\sigma_{2}}$, but also the joint photon-number distribution of these two
beams. This distribution is written as the joint photon statistics matrix
$\mathbf{\sigma}$, where $\sigma_{m,n}$ is the probability of simultaneously
having $m$ photons in mode 1 and $n$ photons in mode 2. We extend Eq.
(\ref{equ:singleTMD}) to relate the probability $P_{m,n}$, of getting outcome
$m$ at detector 1 and outcome $n$ at detector 2, to the joint photon
statistics $\mathbf{\sigma}$,
\begin{equation}
\mathbf{P}=\mathbf{C}_{1}\cdot\mathbf{L}\left(  \eta_{1}\right)
\cdot\mathbf{\sigma}\cdot\mathbf{L}^{\mathrm{T}}\left(  \eta_{2}\right)
\cdot\mathbf{C}_{{2}}^{\mathrm{T}},\label{equ:jointTMDstat}%
\end{equation}
where subscripts indicate the relevant detector and $\mathbf{A}^{\mathrm{T}}$
is the transpose of $\mathbf{A}$. Joint photon statistics are a measure of
photon-number correlations in beams $1$ and $2$, and are thus sensitive to loss
as discussed in the introduction. We use this description of PNRDs to
generalize the Klyshko calibration.

\section{Generalizing the Klyshko method}

\label{genklyshko}

The assumption that only a single photon pair is generated at a time in the Klyshko efficiency-estimation scheme is only valid for very low SPDC pump powers. We can determine in what manner the Klyshko scheme breaks down when more than one
pair is produced. Simply assuming an additional contribution of 
pairs $\sigma_{2,2}$ to the pair rate $\sigma_{1,1}$ changes the
efficiency estimate of Eq. (\ref{equ:originalKlyshko}) to
\begin{equation}
\tilde{\eta}_{s}=\frac{\tilde{R_{c}}}{\tilde{R_{i}}}=\frac{\sigma_{1,1}+\left(  \eta_{s}%
\eta_{i}-2\eta_{s}-2\eta_{i}+4\right)  \sigma_{2,2}}{\sigma_{1,1}+\left(
2-\eta_{i}\right)\sigma_{2,2}  }\eta_{s}. \label{equ:badklyshko}%
\end{equation}
This overestimates the efficiency for $\sigma_{2,2}\neq0$. Similarly, sensitivity to the detailed photon statistics of the input states acts to degrade the accuracy of the efficiency estimate. Let
us consider the effect of background light (e.g. fluorescence from the optical
elements and detector dark counts) on the efficiency estimation of Eq. (\ref{equ:originalKlyshko}). Background photons are uncorrelated between the
signal and idler beams, increasing the singles rate, and thus increasing $R_{i}.$ This will
make the estimated efficiency of the detector $\tilde{\eta}_{s}$ lower than
the actual efficiency $\eta_{s}$.

To generalize the Klyshko scheme to PNRDs and avoid the
above limitations we need to determine the true state generated by SPDC photon-pair
sources. Ideally, these sources produce a `two-mode vacuum squeezed state'
\begin{equation}
\left\vert \Psi\right\rangle =\sqrt{1-|\lambda|^{2}}\sum_{n}\lambda
^{n}\left\vert n\right\rangle _{s}\left\vert n\right\rangle _{i},
\label{equ:thermal}%
\end{equation}
where $\lambda$ is proportional to the pump beam energy, and $\left\vert n\right\rangle _{s(i)}$ is an $n$-photon state of the signal (idler) mode. Having only one free
parameter, it is tempting to use this state in the generalized efficiency
estimation. However, this state can only be generated with careful source
design \cite{Mosley08}. Instead, photons are typically generated in many
spectral and spatial modes in the signal and idler beams \cite{Yamamoto2006}. Depending
on the number of modes in the beams, the thermal photon-number distribution in
Eq. (\ref{equ:thermal}) changes continuously to a Poisson distribution \cite{Mandel1959}. Consequently, it would be incorrect to assume the source produces an ideal two-mode vacuum squeezed state. Still, the number of photons remains perfectly
correlated between the two beams. Without access to the number of
generated modes we can make only the following assumption about the
joint photon statistics of the source%
\begin{equation}
\sigma_{m,n}=c_{m}\cdot\delta_{m,n}, \label{equ:photoncorr}%
\end{equation}
where $\left\{  c_{i}\right\}  $ are arbitrary up to a normalization constant and
$\delta_{m,n}$ is the Kronecker delta.

Inserting the joint photon statistics defined by Eq. (\ref{equ:photoncorr})
into Eq. (\ref{equ:jointTMDstat}) we arrive at the basis for our generalized
Klyshko method. Since $\mathbf{C_{1}}$ and $\mathbf{C}_{\mathbf{2}}$ of the
detectors are known the predicted joint outcome probabilities $\mathbf{P}$ are
highly constrained having $N^{2}$ elements uniquely defined by the $N$
parameters in $\left\{  c_{i}\right\}  $ and the two efficiencies $\eta_{1}$
and $\eta_{2}$. Consequently, a measurement of the joint outcome statistics
specifies $\eta_{1}$ and $\eta_{2}$ with a large amount of redundancy; the
efficiencies are overdetermined. In order to correctly incorporate all measured
outcome statistics into the efficiency estimates a numerical
optimization approach is used. We minimize the difference $\mathbf{G}$ between the
measured outcome statistics $\mathbf{R}$ and the predicted outcome statistics
$\mathbf{P,}$ which are determined by $\left\{  c_{i}\right\}  ,\eta_{1}$, and
$\eta_{2}$
\begin{equation}
\mathbf{G}=\mathbf{R}-\mathbf{C_{1}}\cdot\mathbf{L}\left(  \eta_{1}\right)
\cdot\mathbf{\sigma}\cdot\mathbf{L}^{\mathrm{T}}\left(  \eta_{2}\right)
\cdot\mathbf{C}_{\mathbf{2}}^{\mathrm{T}}.
\end{equation}
This is done by minimizing the Frobenius norm$\ F=\left\{ \mathrm{Tr}\left[\left(
\mathbf{G}\right)^{2}\right ]\right \}^{1/2}  $ to find the optimal $\eta_{1}$ and
$\eta_{2}$ -- our estimates of the PNRD efficiencies. Using the
Frobenius norm makes this a least-squares optimization problem over $\left\{
c_{i}\right\}  ,\eta_{1}$ and $\eta_{2},$ where $0\leq\eta_{i}\leq1$.
However, this method of efficiency estimation is amenable to other
optimization techniques such as maximum-entropy or maximum-likelihood
estimation. We use the Matlab\textregistered{} function `lsqnonneg' (with the
constraints $\sigma_{m,m}\geq0$). %
\begin{figure}
[ptb]
\begin{center}
\includegraphics[
height=3.4324in,
width=4.5455in
]{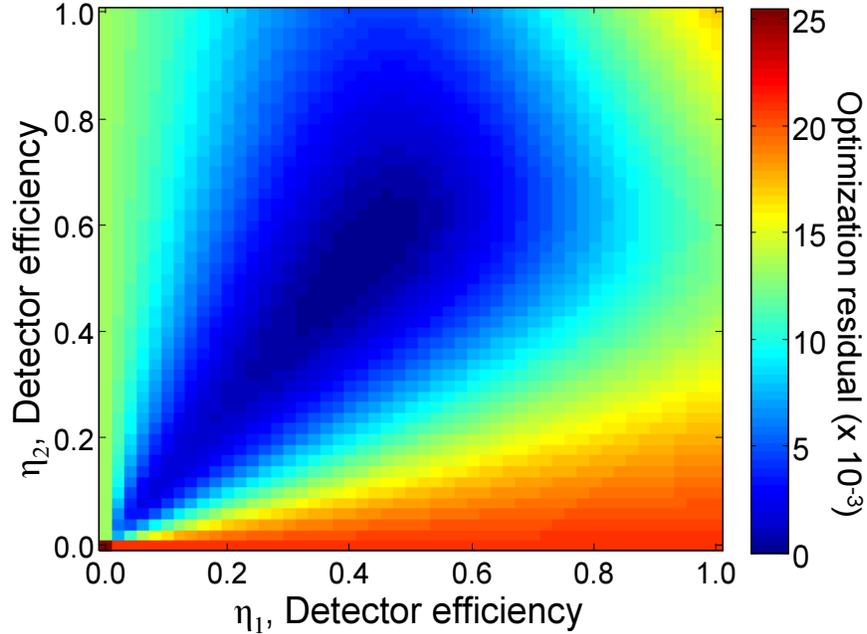}%
\caption{A typical optimization residual ($F$) for simulated joint outcome
statistics as a function of the two PNRD efficiencies $\eta_{1}$ and $\eta
_{2}.$ There is only one minimum, suggesting that the problem is convex.}%
\label{fig:lossspace}%
\end{center}
\end{figure}

This efficiency estimation is similar to the optimizations performed in state or process
tomography, which are known to be convex problems (i.e. there are no local
minima) \cite{Kosut2004}. However, we are now estimating parameters in both 
our state $\hat{\rho}$ and the POVM set $\left\{\hat{\Pi}_{i}\right\}.$
Because this is not guaranteed to be convex \cite{Kosut2004}, we simulated a variety of measured statistics to test for a single minimum. 
Using $\mathbf{C_{1}}$ and $\mathbf{C}_{\mathbf{2}}$ for
the PNRDs in our experiment (TMDs), with several sets of photon
statistics $\left\{  c_{i}\right\}  ,$ and a range of efficiencies $\eta_{1}$
and $\eta_{2},$ we simulated various outcome statistics. In all cases, the
optimization reproduced the correct efficiencies and only a single minimum was observed. Figure \ref{fig:lossspace} shows the result of a typical simulation displaying
a global minimum in the optimization residual (the minimum value of $F$ for a given pair of efficiencies $\eta_{1}$
and $\eta_{2}$). This is a good indication that efficiency estimation
is a convex problem.

\section{Experimental setup}

To experimentally demonstrate and test our efficiency estimation method, time-multiplexed detectors (one possible realization of a PNRD) were employed. In a TMD, the input optical state is contained in a pulsed wavepacket mode. The pulse is split into two spatial and several temporal modes by a network of
fiber beam splitters and then registered using two APDs \cite{Achilles04}. APDs produce largely the same response for one or more incident photons. The TMD overcomes this binary response by making it
likely that photons in the input pulse separate into distinct modes and
are thus individually registered by the APDs \cite{Achilles04}. The TMD is a well-developed technology, which makes this an ideal detector to test our approach to detector efficiency estimation. The convolution matrix $\mathbf{C}$ for this detection scheme is calculated from a classical model of the detector using the fiber splitting ratios \cite{Achilles04}, and is also reconstructed using detector tomography
\cite{Coldenstrodt-Ronge2008,Lundeen2008}. Loss effects in TMDs have also been thoroughly investigated \cite{Achilles06}. In our experiments, the TMDs have four time bins in each of two spatial modes, giving resolution of up to eight photons, with a possible input pulse repetition rate of up to $1$ MHz. Field-Programmable-Gate-Array (FPGA) electronics are used to time gate the APD signals with a window of $4$ ns, which
significantly cuts background rates. The joint count statistics $\mathbf{R}$
are accumulated by the electronics and transferred to a computer for data analysis.%
\begin{figure}
[ptb]
\begin{center}
\includegraphics[
height=3.2379in,
width=4.5455in
]%
{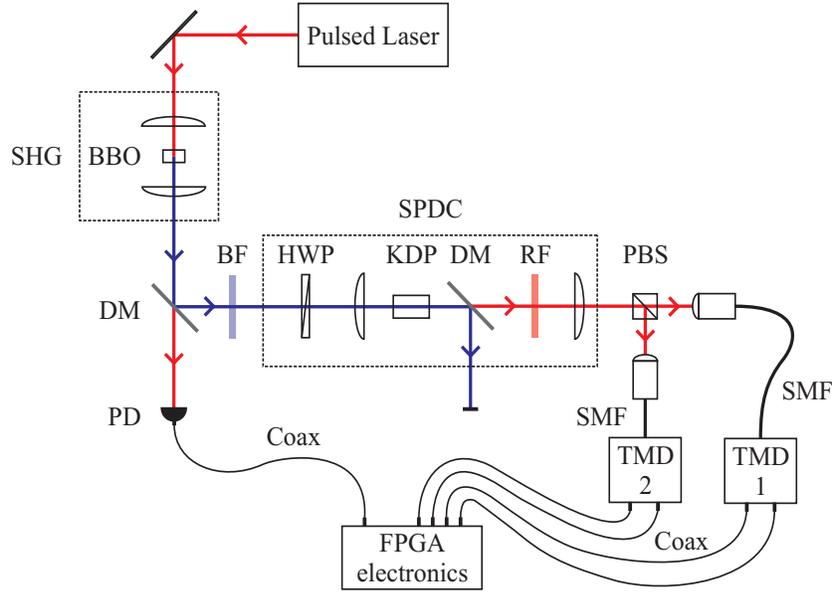}%
\caption{The experimental setup. A KDP crystal two-beam source is pumped by
an amplified Ti:Sapph laser. The two generated beams propagate collinearly and
are orthogonally polarized. Each beam is measured by a time-multiplexed photon
number resolving detector. Details are given in the text.}%
\label{fig:setup}%
\end{center}
\end{figure}

The experimental setup is centered on a nearly-two-mode SPDC
source \cite{Mosley08} as depicted in Fig. \ref{fig:setup}. The twin beam state produced by SPDC in a potassium dihydrogen phosphate (KDP) nonlinear crystal consists of two collinear beams with orthogonal polarizations. The pulsed pump (415 nm central wavelength) driving the SPDC is a frequency-doubled amplified Ti:Sapphire laser operating with a 250 kHz repetition rate. A pick-off beam is sent to a fast photodiode (PD) that is used to trigger the detection electronics. Dichroic mirrors (DM) and a red-pass color glass filter (RF) are used to separate the blue pump from the near-infrared (830 nm central wavelength) SPDC light. A polarizing beam splitter (PBS) is used to separate and direct the two co-propagating downconversion beams into separate single-mode fibers (SMFs) connected to the time-multiplexed detectors (TMD1 and TMD2). The joint statistics $\mathbf{R}$ of the two TMDs are recorded for a range of pump powers between 1 and 55 mW in order to estimate the two TMD efficiencies at each power.
To examine the spectral response of the detector efficiency, one could tune the wavelength of the SPDC source by adjusting the pump wavelength and the crystal orientation.

\section{Estimated efficiencies}

\begin{figure}
[ptb]
\begin{center}
\includegraphics[
height=3.3857in,
width=4.5446in
]%
{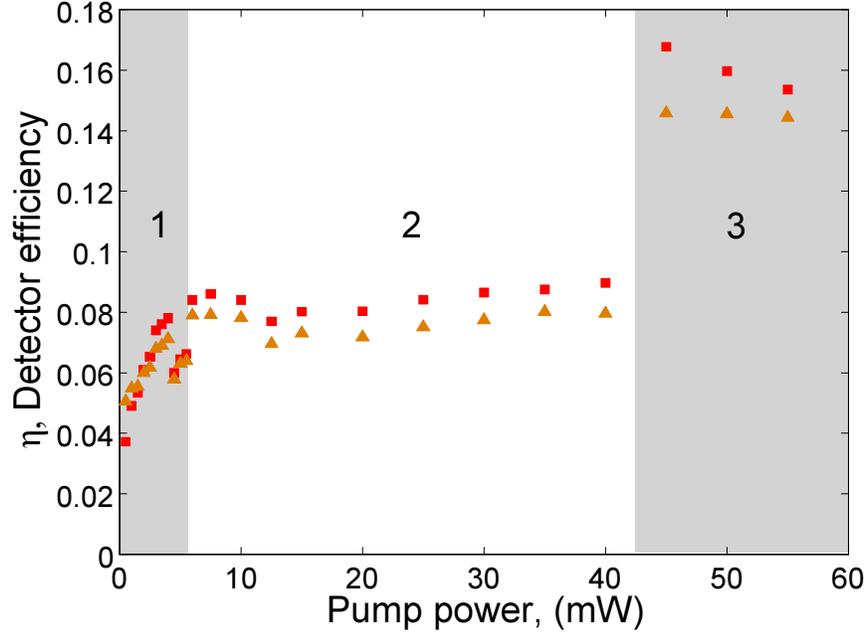}%
\caption{The estimated detector efficiencies for TMD$_{1}$ (\textcolor{red}{$\blacksquare$})
and TMD$_{2}$ (\textcolor{orange}{$\blacktriangle$}) as a function of the average SHG power
pumping the SPDC. Three distinct regimes (regions 1, 2, and 3) are indicated
by shading.}%
\label{fig:efficiencies}%
\end{center}
\end{figure}

For each pump power we determine the optimum detector efficiencies that are consistent with the measured statistics at both PNRDs, shown in Fig. \ref{fig:efficiencies}. Three different regimes are observed.
At low powers (up to 6\,mW) the estimated efficiency increases with power.
Between powers of 6 and 40\,mW the estimates appear
constant. At 40 mW there is a sudden jump in the estimated efficiency (approximately twice the previous value); above this power the estimates remain constant. Continued investigation revealed that
the second-harmonic generation (SHG) process qualitatively changed its behavior at 40\,mW: the increased pump power induced unwanted higher-order nonlinear effects, resulting in the generation of additional frequency components other than the second harmonic and a change in the spatial mode structure. This changed both the transmission of the short wave pass filter (DM and BF) and the efficiency of the fiber coupling into our detectors. Since this behavior is outside the scope of our investigation we omit this data from further discussion.%

The TMD efficiency should be independent of the average photon
number of the state and thus the pump power. This is true in the second region of Fig. \ref{fig:efficiencies}
but not the first. By reconstructing the joint photon-number distribution of the input state ($\sigma_{m,n}$) using the estimated efficiencies, one can gain insight into the estimation accuracy of the detector efficiency.
\begin{figure}
[ptb]
\begin{center}
\includegraphics[
height=2.1in,
width=4.5455in
]%
{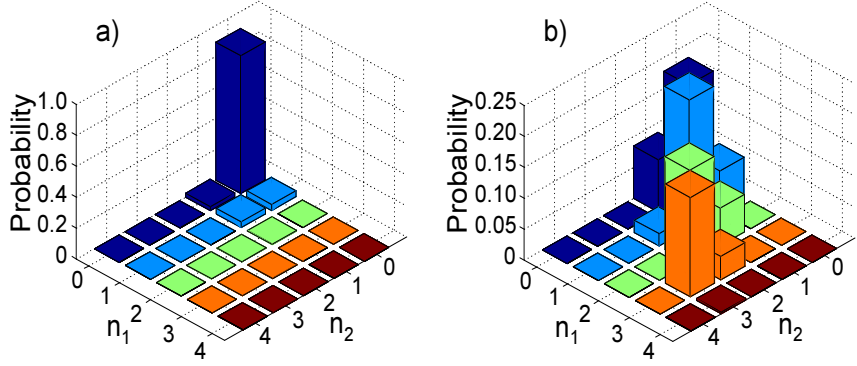}%
\caption{The reconstructed photon statistics $\mathbf{\sigma}$ for $n_{1}$
and $n_{2}$ photons in beam $1$ and $2$, respectively, measured at a pump
power of (a) 1.5 mW and (b) 30 mW. The presence of significant off-diagonal
elements indicate that the input state is corrupted by background.}%
\label{fig:states}%
\end{center}
\end{figure}
This serves as a partial check for our assumption that the number of
photons in the two beams is equal. In Fig. \ref{fig:states}, we show the
reconstructed joint photon statistics for two pump powers in regimes 1 and 2.
In the second regime, the photon-number distribution is largely diagonal: only $10\%$ of the incident photons arrive without a partner in the other beam. In
contrast, the state in the low power regime has significant off-diagonal
components, with $43\%$ of the photons arriving alone. This suggests that at low powers the reference state is corrupted by background photons, possibly fluorescence from optics in the pump beam path, pump photons leaking through our filters, or scattered pump photons penetrating our fiber coatings. Contributions from dark counts are expected to be negligible, since the specified dark count rates of our detectors are significantly lower than these other effects. In the next section we attempt to remove this background in order to better estimate the efficiency.

\section{Compensation for background light}

We investigate two methods of dealing with background light. In the first,
a parameterized background contribution is incorporated into the efficiency
estimation procedure. The second attempts to measure the outcome distributions
due to the background alone and then subtract these from the efficiency data.

\begin{figure}
[htb]
\begin{center}
\includegraphics[
height=3.45114in,
width=4.09095in
]%
{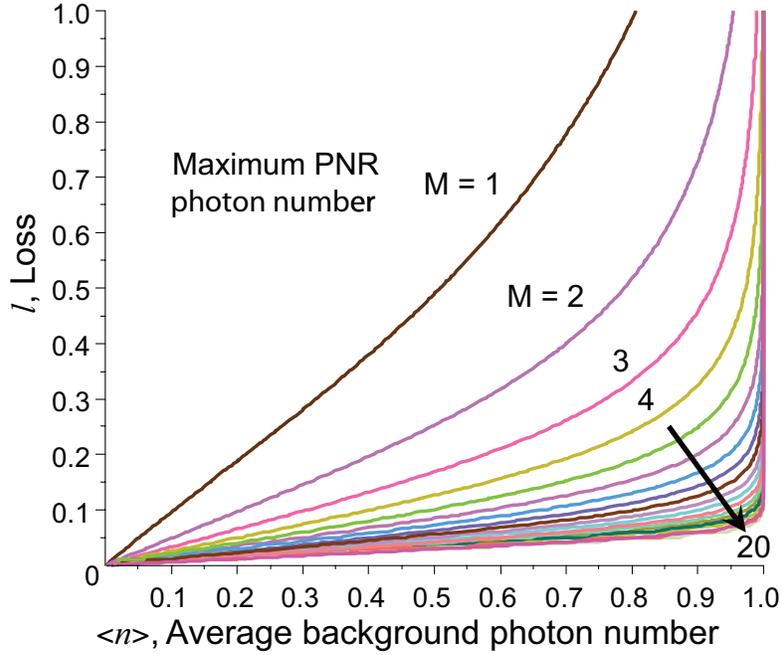}%
\caption{The loss $l$ in beam 1 of a twin-beam state $\mathbf{\sigma}^{B}$
that results in the same joint outcome statistics as the addition of a
Poissonian background, with average photon number $\left\langle n_{B}%
\right\rangle ,$ to beam 2 of another twin-beam state, $\mathbf{\sigma}^{A},$
for some $\mathbf{\sigma}^{A}$ and $\mathbf{\sigma}^{B}$. $M$ is the photon
number range of PNRDs in beams 1 and 2.}%
\label{fig:lossvsbkgrnd}%
\end{center}
\end{figure}

As pointed out in Section 3, the efficiency estimation is greatly
overdetermined. One expects that a modeled background could be added
to our input state model, Eq. (\ref{equ:photoncorr}), without jeopardizing the convergence of the
optimization. This background would be entirely uncorrelated,
possibly making its contribution to the joint outcome statistics easily
distinguishable from the SPDC contribution. We model the
photon-number statistics of the background in each beam by a Poisson photon-number distribution
$d(n)=\alpha^{n}\exp(-\alpha)/n!$ \cite{footnote}, which is fixed by a single parameter -- the
average background photon number, $\alpha=\left\langle n_{B}\right\rangle$.
Consequently, only two additional parameters (one for each beam) enter into the efficiency estimation, keeping it overdetermined. Unfortunately, we theoretically found that the loss in one
beam transforms the outcome statistics in a manner similar to background in the
other beam. Thus, the problem is no longer convex; there is a set of equally optimal points $\{(\left\langle n_{B}\right\rangle ,\eta)\}$. To show this we compare two different two-mode number-correlated states $\sigma^{A(B)}$, similar to the state in Eq. (\ref{equ:photoncorr}), that undergo the addition of uncorrelated background light and loss respectively. The addition of background to beam 1 of state $\sigma^A$ is given  by the convolution of the Poisson distribution with input state, $\mathbf{\sigma}^{A}$ (defined by arbitrary $\left\{c^{A}\right\}$).
The elements of the background-added joint photon-number statistics $\mathbf{\tilde{\sigma}}^{A}$ are given by the sum of all possible ways to add photons from the Poisson background, $d$, to the first beam of the initial state, $\mathbf{\sigma}^{A}$, that add up to a particular number of photons in the final state, $\mathbf{\tilde{\sigma}}^{A}$
\begin{equation}
\mathbf{\tilde{\sigma}}^{A}_{k,l}=\sum_{m=0}^{k}d\left(m\right)\cdot\mathbf{\sigma}^{A}_{k-m,l}.
\label{eq:bgexplain}
\end{equation}
This can be written in terms of a matrix product
\begin{equation}
\mathbf{\tilde{\sigma}}^{A}=D\left(\left\langle n_{B}\right\rangle \right)  \cdot\mathbf{\sigma}^{A},
\label{eq:bgexplain1}
\end{equation}
where the elements of $D\left(\left\langle n_{B}\right\rangle \right)$, $D_{m,n}=d(n-m)$ are the probabilities of having an additional $n-m$ photons from the Poissonian background. To compare this background-added case with the situation in which there is no background, but there is loss, we assume a loss $l=(1-\eta)$ in beam 2 of the second state $\mathbf{\sigma}^{B}$ (defined by arbitrary $\left\{c^{B}\right\}  $). We then attempt to show that for some $\mathbf{\sigma}%
^{A}$ and $\mathbf{\sigma}^{B}$ (each with perfectly correlated photon
statistics) the two resulting states are equal, that is%
\begin{equation}
D\left(  \left\langle n_{B}\right\rangle \right)  \cdot\mathbf{\sigma}%
^{A}=\mathbf{\sigma}^{B}\cdot\mathbf{L}^{\mathrm{T}}\left(  \eta\right).
\label{eq:background}%
\end{equation}
Elimination of $\left\{c^{B}\right\}$ and $\left\{c^{A}\right\}$ from the resulting
equations is facilitated using a computer algebra program, and we find that there is a range of $l$ and $\left\langle
n_{B}\right\rangle \leq1$ that solve Eq. (\ref{eq:background}), indicating that
it is not possible to simultaneously fit for background and efficiency. This emphasizes the fact that one cannot distinguish between loss and Poisson background for the number-correlated states. Figure \ref{fig:lossvsbkgrnd} shows $l$ as a function of $\left\langle n_{B}\right\rangle $ for PNRDs with a maximum photon number of $M=1$
to $20$.
The standard Klyshko case corresponds to $M=1$. Note that as $M$
becomes large the loss curve converges to line of slope $0.068$ suggesting
that even PNRDs with an infinite photon number range would not allow
one to distinguish loss and background. This suggests that another approach to background light should be used.

Another approach to address the background contribution to the efficiency estimation attempts to subtract an independently measured background contribution from the click statistics. For each pump power the pump polarization is rotated by 90 degrees, extinguishing the SPDC
and allowing the measurement of the joint outcome statistics due to background
alone. Generally, the statistics of two concurrent but independent processes
is the convolution of the statistics of the two processes. However, the
measured outcome probabilities $\mathbf{P}_{M}$ of both light sources combined is not a simple convolution of the background outcome probabilities $\mathbf{P}_{B}$ and the twin-beam outcome
probabilities $\mathbf{P}_{S}$ 
\begin{equation}
\mathbf{P}_{M}\neq\mathbf{P}_{S}\ast\mathbf{P}_{B}.\label{eq:clickdneconv}%
\end{equation}
\begin{figure}
[bht]
\begin{center}
\includegraphics[
height=2.15in,
width=4.5455in
]%
{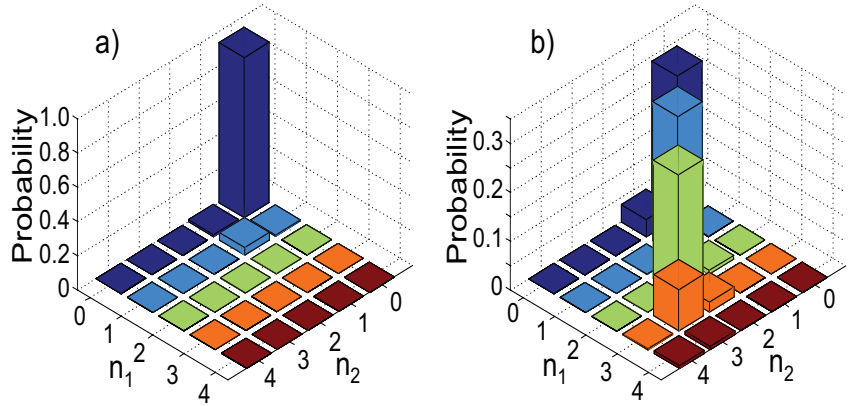}%
\caption{The reconstructed photon statistics $\mathbf{\sigma}$ for $n_{1}$
and $n_{2}$ photons in beam $1$ and $2$, respectively, after subtracting an
independently measured background. For pump powers (a) 1.5 mW and (b) 30
mW, as in Fig. \ref{fig:states}. The reduction of significant
off-diagonal elements indicates that the background subtraction method works.}%
\label{fig:bgsubstates}%
\end{center}
\end{figure}

This is due to the fact that there is an effective interaction between the background and twin-beam signal detection events due to the strong detection nonlinearity of the APDs (i.e. the detectors saturate at one
photon).
However this saturation effect can be eliminated by first applying the inverse
of the $\mathbf{C}$ matrices to the measured statistics
\begin{align}
\mathbf{P}_{M}^{\prime} &  =\mathbf{C}_{1}^{-1}\mathbf{P}_{M}(\mathbf{C}%
_{2}^{\mathrm{T}})^{-1},\label{eq:calcreduced0}\\
\mathbf{P}_{B}^{\prime} &  =\mathbf{C}_{1}^{-1}\mathbf{P}_{B}(\mathbf{C}%
_{2}^{\mathrm{T}})^{-1},\label{eq:calcreduced1}%
\end{align}
which gives the estimated photon-number statistics prior to the mode multiplexing. These non-mode-multiplexed statistics can be assumed independent so that,
\[
\mathbf{P}_{M}^{\prime}=\left[  \mathbf{C}_{1}^{-1}\mathbf{P}_{S}%
(\mathbf{C}_{2}^{\mathrm{T}})^{-1}\right]  \ast\mathbf{P}_{B}^{\prime}.
\]
We then use the convolution theorem to find
\begin{equation}
\mathbf{P}_{S}\ =\mathbf{C}_{1}\mathcal{F}^{-1}\left\{  \frac
{\mathcal{F}\{\mathbf{P}_{M}^{\prime}\}}{\mathcal{F}\{\mathbf{P}_{B}^{\prime
}\}}\right\}  \mathbf{C}_{2}^{\mathrm{T}},\label{eq:deconv}%
\end{equation}
where $\mathcal{F}$ indicates the Fourier transform, and the matrix division
is element by element. Using $\mathbf{P}_{S}$ we estimate the efficiency as
previously by the method in Section 3.%

\begin{figure}
[bth]
\begin{center}
\includegraphics[
height=3.2776in,
width=4.5455in
]%
{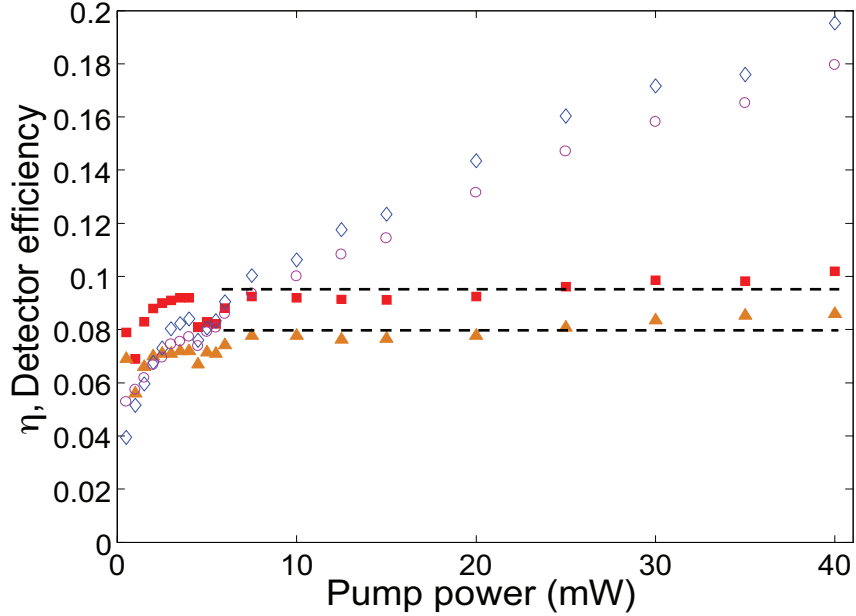}%
\caption{The estimated detector efficiencies for TMD$_{1}$ (\textcolor{red}{$\blacksquare$})
and TMD$_{2}$ (\textcolor{orange}{$\blacktriangle$}) determined from background subtracted outcome statistics,
plotted as a function of the average pump power. Also plotted is the Klyshko
efficiencies that would have been estimated for single photon detector 1
(\textcolor{blue}{$\lozenge$}) and 2 (\textcolor{purple}{$\bigcirc$}). The standard Klyshko method overestimates the
efficiencies for high powers. The dotted lines indicate the average efficiencies of the two PNRDs.}%
\label{fig:bgsubeffs}%
\end{center}
\end{figure}

To test the accuracy of this background subtraction method we reconstruct the joint
statistics at the same powers as in Fig. \ref{fig:efficiencies}. We find that
in both cases the off-diagonal components are significantly reduced. In the low power regime, now only $16\%$ of incident photons are not part of a pair, and at higher power this becomes just $4\%$. With the background subtracted, the estimated efficiencies are plotted in Fig. \ref{fig:bgsubeffs} and are now in better agreement with the expected constant detector efficiency through the first two regions. However, the estimates still drop off as the power goes very low. Although we measure the background, we do not do so \textit{in situ}; we need to change the apparatus by rotating a half wave plate (HWP). Thus, we are not guaranteed that this is equal to the background present during the efficiency estimation. Furthermore, errors in background measurements are more significant at low powers as background then forms a larger component of the outcome statistics.

Also plotted in Fig. \ref{fig:bgsubeffs} is the Klyshko efficiency. In
contrast to the increased dynamic range of our method the standard Klyshko
efficiency increases with pump power, evidence that higher photon numbers in
the input beams distort the estimated efficiency.

The average efficiency across the second region was found to be $9.4\%\pm0.4\%$
for detector 1 and $8.0\%\pm0.4\%$ for detector 2, where the errors are the
standard deviations.
These relatively low efficiencies are only partly due to the
quantum efficiency of the avalanche photodiodes themselves, which is specified to
be $60\%\pm5\%$ at our wavelength. Bulk crystal SPDC sources ordinarily emit into
many spatial modes, which makes coupling into a single-mode
fiber difficult and inefficient. Typical coupling efficiencies are less than $30\%$ \cite{Mosley08a}. 

\begin{figure}
[ptb]
\begin{center}
\includegraphics[
height=3.6045in,
width=4.5455in
]%
{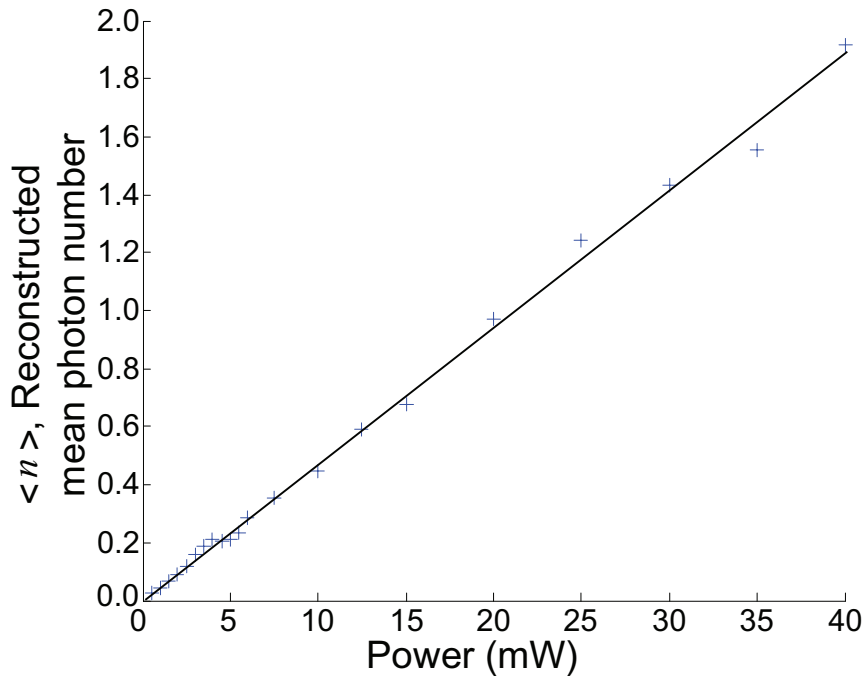}%
\caption{The reconstructed average photon number ($+$) as a function of the
pump power. SPDC theory predicts a linear relationship. The line is a linear
fit.}%
\label{fig:linearn}%
\end{center}
\end{figure}

As a final check of the reconstructed photon statistics, we plotted the average reconstructed photon
number as a function of pump power in Fig. \ref{fig:linearn}. The relationship is linear, as expected when taking into account the higher dynamic range of a TMD in comparison to standard APDs \cite{Coldenstrodt-Ronge2007}.%

\section{Conclusion}

Despite being one of the seven base SI physical quantities, the working
standards for luminous intensity have a relative accuracy of only $0.5\%$ \cite{Ohno1997}, compared to $10^{-12}$ and better for the second \cite{BIPM}. Relying on a light beam of a known intensity, the efficiency calibration of detectors is similarly limited. Quantum states of light give us the opportunity to bypass the working standards and calibrate
detectors directly. Indeed, the photon has been suggested as an alternative to the candela as the
definition of luminous intensity \cite{Cheung2007}. We use twin-beam states and their
perfect photon-number correlations to measure the efficiency of a photon-number-resolving detector. This type of state can be produced at a wide range
of wavelengths from many different types of source including nonlinear
crystals, optical fibers, periodically poled waveguides, and atomic gases. This
method has the advantage that it only assumes perfect photon-number correlation and does not assume the state has a specific photon-number distribution, nor even that it is pure. Despite these seemingly detrimental assumptions, the efficiency estimation presented has a large amount of redundancy leading to a relatively small
absolute error of $0.4\%$. We show that this measurement is independent of
the average photon number of the state, unlike the Klyshko method, making it more
widely applicable. In particular, it is ideal for characterizing photon-number
resolving detectors, and for use with bright reference states. PNR detectors are 
undergoing rapid development via a number
of competing technologies. These detectors will play an important role in precision optical 
measurements and optical quantum information protocols where photon-number
resolution is necessary for large algorithms. Future development of direct
efficiency calibration should focus on the issue of background, which can
corrupt the state and thus the calibration.

\section*{Acknowledgments}
The disclosure in this paper is the subject of a UK patent application (Ref: 3960/rr). Please contact Isis Innovation, the Technology Transfer arm of the University of Oxford, for licensing enquires (innovation@isis.ox.ac.uk).

This work has been supported by the European Commission under the Integrated
Project Qubit Applications (QAP) funded by the IST directorate Contract
Number 015848, the EPSRC grant EP/C546237/1, the QIP-IRC project and the Royal Society. HCR
has been supported by the European Commission under the Marie Curie Program
and by the Heinz-Durr Stipendienprogamm of the Studienstiftung des deutschen Volkes.

\end{document}